\documentclass[11pt]{article}
\usepackage{hyperref}
\usepackage{cite}
\usepackage{graphicx}
\usepackage{subfigure}
\usepackage[top=3cm, bottom=3cm, left=1.5cm, right=1.5cm]{geometry}
\usepackage{amsmath,amsfonts,amssymb,mathtools}
\usepackage{graphicx}
\usepackage{float}
\usepackage{breqn}
\usepackage{multicol}
\usepackage{multirow}
\usepackage{perpage}
\usepackage{booktabs}
\usepackage{xcolor}
\usepackage{array,makecell, cellspace}
\date{}

\begin{document}
	
	\setcounter{page}{1}
	\pagestyle{plain}

	\begin{center}
		\Large{\bf  Higgs Inflation in Unimodular Gravity}\\
		\small \vspace{1cm} {\bf Kourosh
			Nozari\footnote{knozari@umz.ac.ir(Corresponding Author)}\quad and \quad{\bf Manda
				Malekpour\footnote{m.malekpour@stu.umz.ac.ir}}}
		\\
		\vspace{0.25cm}
		Department of Theoretical Physics, Faculty of Science,\\
		University of Mazandaran,\\
		P. O. Box 47416-95447, Babolsar, Iran
	\end{center}
\begin{abstract}
	The discovery of Higgs mechanism within the context of spontaneous symmetry breaking has offered a new perspective on the early time cosmic inflation and also on the relationship between elementary particles and dark energy, believed to drive the universe's accelerating expansion. We suggest an inflation scenario driven by the Higgs boson within the framework of unimodular gravity, where the Higgs field acts as the inflaton and has a significant non-minimal coupling to the gravity. We present a detailed analysis of the problem in the Jordan and then Einstein frame for a unimodular Higgs inflation, followed by a comparison of our findings with the Cosmic Microwave Background observations made by the Planck Collaboration and other joint data sets. Therefore, new constraints are imposed on the non-minimal coupling parameter, $\xi$, by determining the magnitudes required for effective cosmic inflation. We demonstrate that a substantial non-minimal coupling of order $\xi\sim 10^{2}-10^{4}$ is required for this model to match with the observed primordial spectrum.\\
{\bf PACS}: 04.50.Kd, 98.80.-k, 98.80.Bp, 47.10.Fg\\
{\bf Keywords}: Higgs Inflation, Unimodular Gravity, Non-Minimal Coupling, Conformal Transformation, Cosmological Constant.\\
\end{abstract}

\newpage
\section{Introduction}
In early 1980, the inflationary scenario as an evolutionary idea, was suggested to solve the cosmological puzzles. In 1981, Alan Guth attempted to clarify the inception conditions of the Big Bang cosmology \cite{guth1981inflationary}. In contemporary cosmology, one of the main concerns is the phenomenon of \textit{\textquotedblleft cosmological inflation\textquotedblright}, which has attracted significant attention in recent years \cite{Senatore:2016aui, baumann2018tasi, Odintsov:2023weg}. An exponential expansion in the early universe, cosmic inflation, can explain temperature fluctuations and anisotropies of the Cosmic Microwave Background (CMB) \cite{Garcia-Bellido:2011kqb, Giovannini:2022vha}. Inflation addresses issues in standard Big Bang cosmology like horizon, flatness, baryon asymmetry, and generates density fluctuations for structure formation observed today \cite{Panotopoulos:2014hwa}. In a so-called slow-roll inflation, a canonical scalar field, \textquotedblleft inflaton,\textquotedblright rolls slowly towards the minimum of its potential \cite{starobinsky1979relict, A.D. Linde, Martin:2013tda, wang2014inflation, senatore2016lectures, martin2016have, Malekpour:2023lsf}. The distribution of stuff in the universe \cite{Kafatos:1996uta} and the measurements of CMB fluctuations constrain the properties of the inflaton \cite{Rio02}. However, the origin of the inflaton is closely linked to the naturalness of inflation \cite{DeMartini:2017izr}. The fundamental question of what's the true nature of the inflaton is still an open issue \cite{Lebedev:2011aq,SanchezLopez:2023ixx}.

To understand the condition of the Big Bang theory, it is necessary to have a valid theory that covers up the inflation scale \cite{Bezrukov:2008cq}. There are various theories proposed to explain physics beyond the Standard Model (SM). Physics theories propose ideas such as Grand Unification, low-energy supersymmetry, tiny, huge, or infinite Extra Dimensions, etc. Meanwhile, an intriguing inflationary model, which is often essential, incorporates the Non-Minimal Coupling (NMC) between the scalar field and the Ricci scalar \cite{Faraoni:2000wk}. It appears impossible that $\xi$ (the NMC parameter) to be discarded or neglected in majority of inflationary theories \cite{Makino:1991sg,M.De,faraoni1996nonminimal,faraoni1998conformal,faraoni2000inflation,park2008inflation,hertzberg2010inflation,qiu2012reconstruction,shaposhnikov2013cosmology,chiba2015consistency,Myrzakul:2015gya,bostan2019quartic}.

In recent years, it has been proven that the SM itself can trigger inflation. In the past, researchers have investigated how the SM can be non-minimally coupled to gravity \cite{Bezrukov:2007ep, Rubio:2018ogq}, represented as $\xi |\phi|^2 R$, resulting in inflation. The Higgs potential exhibits an essentially flat area in the vast field area, potentially resulting in successful inflationary scenarios \cite{Bezrukov:2009db}. This result was achieved by using a tree-level approximation of the theory, as it was argued that its radiative corrections were significantly diminished by a high NMC constant value $\xi$. Note that no additional particles are needed beyond those in the electroweak theory to explain inflation \cite{Barvinsky:2008ia}.

In view of the previous arguments, the final piece component of the SM was completed with the Higgs field $\phi$, which was found at CERN's Large Hadron Collider (LHC) \cite{DeSimone:2008ei, ATLAS:2012yve, CMS:2012qbp} in 2012, paving a new era in particle physics. Thus, cosmology and particle physics have grown closer in the past few decades \cite{Ghoshal:2024ycp}. In particular, the inflaton, when non-minimally coupled to gravity, may potentially be recognized as the Higgs boson within the SM \cite{fakir1990improvement, kaiser1995primordial, Komatsu:1997hv, Komatsu:1999mt, Tsujikawa:2004my}.
The first attempts to describe inflation as a Higgs-like scalar field $\phi$ interacting with itself that is minimally coupled to gravity required an incredibly tiny self-coupling constant $\lambda_{0} \sim 10^{-13}$ of its potential, $V (\phi) = {\lambda_{0}}{\phi}^4/4$ \cite{Pich:2007vu,Barvinsky:2008ia}. At first, this scenario was considered highly unpleasant in terms of reheating after inflation; however, the current observational data actually rule out this straightforward model \cite{linde1983chaotic,akrami2020planck}.
The Higgs potential, a fundamental component of the SM, contributes significantly to the development of the early universe \cite{Kamada:2013bia}. In the presence of $ \xi $, inflation happens with $\xi \phi^{2}/M_{P}^{2} \gg1$, and it has been observed that including the NMC term $\xi \phi^{\dag} \phi R$ with a large coupling constant $\xi$ to the Einstein term solves the problem of small $ \lambda_{0} $ \cite{Bezrukov:2010jz}. This implies that for $\lambda_{0} $ close to the order of 1, $|\xi|\sim 10^{4}$ is sufficiently large\footnote{An unrealistically high level of primordial inhomogeneities is produced by minimal coupling to a Ricci scalar with $\xi=0$ for a quartic Higgs self-interaction \cite{Linde:1983gd, Bezrukov:2010jz}.}, although lower values could be achievable in very special cases, such as the potential for inflation to occur near the critical point \cite{Bezrukov:2014bra ,Saltas:2015vsc}. In contrast, the unitary problem can be resolved through the incorporation of non-minimal derivative couplings \cite{germani2010new,germani2014self}. In other words, the non-minimal derivative coupling inflation model offers an alternative approach to realizing the Higgs model without the need to introduce an additional degree of freedom. This is a significant advantage. Within the realm of particle physics \cite{Mishra:2018dtg}, the NMC constant $\xi $ represents the only remaining free coupling parameter that can be constrained through cosmological observations, while the quartic coupling constant $\lambda_{0} $ is predicted by the theoretical equations of the SM.

The \textquotedblleft cosmological constant\textquotedblright problem is a long-standing and fascinating puzzle in theoretical physics, included in the current study as another component \cite{Herrero-Valea:2020xaq}. Einstein used the phrase cosmological constant in General Relativity (GR) to describe the energy density of the vacuum \cite{Leon:2022kwn}, but its value has proven to be a persistent mystery. The issue is a significant difference (ranging from 50 to 120 orders of magnitude) between the predicted high value of the cosmological constant derived from quantum vacuum energy density and the much lower observed value \cite{Piccirilli:2023klw}.
In light of the favorable outcomes of the inflationary model, it is logical to investigate alternatives to the conventional slow-roll inflationary model while maintaining the accuracy of the inflation predictions. In recent decades, a relatively unexplored modification of GR has gained scattered attention, despite being almost as old as the original theory of GR itself. In 1919, Einstein introduced a novel theoretical framework for the first time to deal with the cosmological constant issue, which is called \textquotedblleft\textit{Unimodular Gravity}\textquotedblright (UG) \cite{einstein1952principle}. This idea was later revisited in the 1970s by Anderson and Finkelstein \cite{anderson1971cosmological}, who further developed and refined the concept of unimodular gravity. This notion has been the subject of numerous studies in recent years; see, for instance, Refs. \cite{gao2014cosmological,cho2015unimodular,Nojiri:2015sfd,Nojiri:2016ppu,Nojiri:2016plt,barvinsky2017darkness,barvinsky2019inflation,barvinsky2019dynamics,leon2022inflation, deCesare:2021wmk}. In this case, the cosmological constant appears as a Lagrange multiplier within the theory \cite{weinberg1989cosmological,unruh1989unimodular,Henneaux:1989zc,liddle1993end,sahni2002cosmological}. Thus, the cosmic constant term is unnecessary in unimodular gravity. As a consequence, this can address the issue of fine-tuning the cosmological constant \cite{jain2012testing,jain2012cosmological,Nojiri:2016ygo,Odintsov:2016imq}. The fundamental concept of unimodular gravity revolves around restricting the value of the metric determinant, $\sqrt{-g}=1$. However, as pointed out by Weinberg in Ref. \cite{weinberg1989cosmological}, the issue remains unresolved, and the matter is not entirely resolved.

In a recent work, we have advanced the understanding of Higgs inflation through the application of a novel analytical technique within unimodular gravity \cite{Malekpour:2023lsf}.
In this paper, based on our previous work \cite{Malekpour:2023lsf}, we endeavor to explore an extension of unimodular non-minimal inflation. First of all, we compute the slow-roll parameters in the Jordan frame and then proceed to the Einstein frame. We propose the exploration of Higgs inflation within the framework of UG. We demonstrate how the Higgs field in UG can lead to inflation in the SM without introducing more degrees of freedom. It means that inflation can arise directly from the SM within this unimodular framework. We conclude that the SM Higgs inflation produces confirmed predictions for observational data and the LHC, and provides predictions that align with present observational data. In this investigation, we will examine the scenario with a high NMC parameter, denoted as $\xi\gg1$, though not large enough to significantly affect the Planck mass within the SM framework ($\phi \sim \upsilon$), which means $\sqrt{\xi} \lll 10^{17}$. Therefore, the Planck mass in our configuration is denoted as $M_{P} = 1/\sqrt{8 \pi G_{N}} = 2.44\times 10^{18}$ GeV \cite{Bezrukov:2008ut}. Specifically, the latest Planck2018 data \cite{akrami2020planck,Planck:2018vyg} and the latest BICEP/Keck data \cite{ ade2021improved} have led to well-defined parameters, the scalar spectral index $ n_{s} $ and the tensor-to-scalar ratio $ r $, as follows
\begin{align*}
	& \mathrm{Planck2018}:\quad\quad n_{s}=0.9658\pm 0.0038\,, \qquad  r<0.072\,,\\
	& \mathrm{BICEP/Keck2021}:\quad\quad r<0.036 \,.
\end{align*}
So, we construct a novel unimodular Higgs inflation model, and we confront our results with observations to see the reliability and feasibility of this model.

The paper is structured in the following manner: In Section \ref{sec-2} and in the context of the unimodular constraint, we examine how the SM Higgs field behaves as an inflaton in the Jordan frame. In Sec. \ref{sec-3}, we transform into the conventional Einstein frame, calculate inflationary potential and other important inflation quantities and discuss how Higgs field drives the cosmic inflation in this frame. Finally, in Sec. \ref{sec-4}, we summarize and conclude our study. In our calculations, we set $ c = \hbar=1$, denote the gravitational constant as $ \kappa^{2}\equiv 8 \pi G $, and use a metric signature of (-, +, +, +).\\

\section{Higgs inflation in the Jordan frame}\label{sec-2}
In the unimdular theory, the metric determinant stays constant, with all components being dynamic, given by $ g_{\mu \nu} \delta g^{\mu \nu}= 0$ \cite{Cho:2014taa}. We assume that the metric determinant is restricted as
\begin{equation}\label{eq1}
	\sqrt{-g}=1\,.
\end{equation}
The Higgs boson in GR \cite{Bezrukov:2008ut} includes a NMC term into the action of the SM. Now let's examine how the inflationary paradigm might be affected by unimodular gravity. Specifically, we examine Higgs inflation in light of the unimodular theory and seek for observational status of this scenario. To elucidate our main idea, let us look at the model's action as follows
\begin{equation}\label{eq2}
	S^{(SM)}_{\phi}= S_{\phi}+S_{SM}\,.
\end{equation}
We consider NMC in the UG framework with a scalar field $f\left(\phi\right)$ coupled to the Ricci scalar $ R$ as	
\begin{equation}\label{eq3}
	S_{\phi}= \int d^{4}x \left\lbrace \sqrt{-g} f\left(\phi \right)  R- \lambda \left( \sqrt{-g}-1\right) \right\rbrace\,,
\end{equation}
where $ \lambda$ is the Lagrange multiplier function and the action of the SM Higgs, $S_{SM}$, is given by
\begin{equation}\label{eq4}
S_{SM}=  \int d^{4}x \sqrt{-g} \left\lbrace -\frac{1}{2} g^{\mu \nu} \partial_{\mu}\phi \partial_{\nu}\phi - V\left( \phi\right)\right\rbrace \,.
\end{equation}
We have intentionally omitted the mass term of the field, $\phi $, since it is not important and therefore inconsequential for $ \phi\gg 1 $ \cite{Bezrukov:2010jz}. The corresponding potential of the standard Higgs boson  is
\begin{equation}\label{eq5}
	V\left( \phi\right)= \frac{\gamma}{4} \left(\phi^{2} -\upsilon^{2}\right)^{2}\,,
\end{equation}
where $\gamma $ ($\gamma \ll 1 $) is the Higgs self-coupling. The Higgs vacuum expectation value is as follows \cite{Pareek:2023die}
\begin{equation}\label{eq6}
\upsilon = 246\, \mathrm{GeV} =1.1 \times 10^{-16} M_{P}\,.
\end{equation}
We consider the NMC of the Higgs field with gravitational sector (which is UG here) usually occurring in certain Higgs inflation scenarios. This scale is sufficiently small $ \upsilon^{2}\ll M_{P}^{2}/ \xi$, as the gravitational interaction is governed by the effective Planck mass $M^{2}_{eff}(\upsilon)=M_{P}^{2}-\xi \upsilon^{2}\simeq M_{P}^{2}$\cite{Kaiser:2010ps}; so we have
\begin{equation}\label{eq7}
	 f\left(\phi \right)=\frac{M_{P}^{2}}{2} \left(1+\frac{\xi \phi^2}{M_{P}^{2}}\right)\,,
\end{equation}
and $\xi $ being the non-minimal coupling constant \cite{Mishra:2018dtg}. Let us consider the action \eqref{eq2} to be rewritten finally as
\begin{equation}\label{eq8}
	S_{J}=\int d^{4}x \left\lbrace \sqrt{-g} \left(f\left( \phi \right) R-\frac{1}{2} g^{\mu \nu} \partial_{\mu}\phi \partial_{\nu}\phi - V\left( \phi\right)\right) - \lambda \left(\phi\right)\left( \sqrt{-g}-1\right) \right\rbrace\,,
\end{equation}
where $ \lambda $ is defined as $ \lambda\left(\phi\right)$, and index \textquotedblleft J\textquotedblright denotes for the \textquotedblleft Jordan frame\textquotedblright action. Note that in the above action, we disregard quantum corrections \cite{Rehman:2010es}. We obtain the field equations as follows
\begin{equation}\label{eq9}
	-\frac{1}{2} g_{\mu\nu} f\left(\phi\right) R + f\left(\phi\right)  R_{\mu\nu} +
	\left( \Box g_{\mu\nu}- \nabla_{\mu} \nabla_{\nu} \right) f\left( \phi\right) + \frac{\left(\lambda \left(\phi\right)+V\left( \phi\right) \right)}{2}  g_{\mu\nu} + \frac{1}{4}g_{\mu\nu} \nabla_{\alpha}\phi \nabla^{\alpha}\phi - \frac{1}{2}\nabla_{\mu}\phi \nabla_{\nu}\phi=0 \,.
\end{equation}
Considering a spatially flat topology, the Friedmann-Robertson-Walker (FRW) metric is defined by
\begin{equation}\label{eq11}
	ds^2 = -dt^{2} + a^{2}\left(t\right) \sum_{i=1}^{3} \left( dx^{i}\right)^{2} \,,
\end{equation}
where $a\left( t\right)$ is the scale factor. The metric \eqref{eq11} does not adhere to the unimodular constraint \eqref{eq1}~\cite{Nojiri:2015sfd}. Therefore, we consider a new time variable as
\begin{equation}\label{eq12}
	d\tau = a^{3}\left(t\right) dt\,.
\end{equation}
So, by using equation \eqref{eq12}, the FRW metric \eqref{eq11} can be rewritten as fallows
\begin{equation}\label{eq13}
	ds^{2} = -a^{-6} \left(\tau\right)  d\tau^{2} + a^{2}\left(\tau\right)  \sum_{i=1}^{3} \left(dx^{i}\right)^{2} \,.
\end{equation}
Now, with the unimodular metric of equation \eqref{eq13}, we obtain the curvatures as
\begin{equation}\label{eq14}
	\begin{split}
		R_{\tau\tau}= -3\dot{\mathcal{H}} - 12 \mathcal{H}^2\,,
		\quad R_{ij}= a^8\left(\dot{\mathcal{H}}+ 6\mathcal{H}^2\right)  \delta_{ij}\,,
		\quad R= a^6 \left(6\dot{\mathcal{H}} +30 \mathcal{H}^2\right) \,,
	\end{split}
\end{equation}
where Hubble rate $\mathcal{H} $ is given by $ \mathcal{H}= \frac{1}{a}\frac{da}{d\tau} $. The components of the field equations are obtained using equations \eqref{eq13} and \eqref{eq14} as follows
\begin{equation}\label{eq15}
	3 \mathcal{H}^{2} f\left(\phi\right)+ 3 \mathcal{H} \dot{f}\left(\phi\right) - \left( \frac{\lambda \left(\phi\right)+V\left( \phi\right)}{2} \right) a^{-6}- \frac{1}{2} \dot{\phi^{2}}=0\,,
\end{equation}
\begin{equation}\label{eq16}
	\left(-2 \dot{\mathcal{H}} -9 \mathcal{H}^{2}\right) f\left(\phi\right)-5 \mathcal{H} \dot{f}\left(\phi\right)- \ddot{f}\left(\phi\right)+\left(\frac{\lambda \left(\phi\right)+V\left( \phi\right)}{2}\right) a^{-6}-\frac{1}{2} \dot{\phi^{2}}=0\,.
\end{equation}
Now, through the combination of equations \eqref{eq15} and \eqref{eq16}, we obtain
\begin{equation}\label{eq17}
	\ddot{f}\left(\phi\right)+2\mathcal{H} \dot{f}\left( \phi\right)+ \left(6 \mathcal{H}^{2}  + 2\dot{\mathcal{H}}\right)  f\left(\phi\right)- \dot{\phi}^{2}=0\,,
\end{equation}
which governs the time evolution of $ f \left(\phi \right)$. In light of current understanding, the universe's expansion rate is characterized by a power-law scale factor with constant values of parameters $t_{0}$ and $b$ as
\begin{equation}\label{eq18}
	a\left(t\right)= \left( \frac{t}{t_{0}}\right)^{b} \Longrightarrow H = \frac{b}{t} \,,
\end{equation}
where we set $b=\frac{2}{3(1+w)}$ that $w$ is the equation of state parameter. By substituting this ansatz into equation \eqref{eq12}, we get
\begin{equation}\label{eq19}
	\tau = \frac{t_{0}}{3b+1} \left( \frac{t}{t_{0}}\right)^{3b+1}\,,
\end{equation}
and then, equation \eqref{eq19} can be put into equation \eqref{eq18} to find
\begin{equation}\label{eq20}
	a\left( \tau\right) = \left( \frac{\left(3b+1\right)  \tau}{t_{0}} \right)^{\frac{b}{3b+1}}\,.
\end{equation}
Thus, we have
\begin{equation}\label{eq21}
	a\left(\tau\right)= \left( \frac{\tau}{\tau_{0}}\right)^{q}   \Longrightarrow \mathcal{H}= \frac{q}{\tau}\,,
\end{equation}
where $q$ and $\tau_{0}$ are both constant values defined respectively as $q=\frac{b}{3b+1}$ and $\tau_{0}=\frac{t_{0}}{3b+1}$. As a consequence, the metric \eqref{eq13} is expressed as follows
\begin{equation}
	ds^{2} = -\left( \frac{\tau}{\tau_{0}} \right)^{-6q} d\tau^{2} + \left( \frac{\tau}{\tau_{0}} \right)^{2 q} \sum_{i=1}^3 \left(dx^{i}\right)^{2}\,.
\end{equation}
In this study, a quintessential evolution occurs within a specific range for $\frac{1}{4}\leq q<\frac{1}{3}$~\cite{Nojiri:2015sfd}. As a consequence, $q=\frac{1}{3}$ indicates that we have a universe with a de Sitter phase of evolution, while $q=\frac{2}{9}$ and $q=\frac{1}{5}$ represent dust-dominated and radiation-dominated universes, respectively. In summary, the value of $q$ can be used to classify the universe into different phases of evolution, each characterized by distinct physical properties. The de Sitter phase, dust-dominated phase, and radiation-dominated phase are the three main phases of evolution that the universe is thought to go through, and each one is characterized by a specific value of $q$. By applying the equation \eqref{eq21}, we obtain
\begin{equation}
	R= \frac{-6q +30 q^{2}}{\tau_{0}^{2}} \left( \frac{\tau}{\tau_{0}}\right)^{6q- 2}\,.
\end{equation}
Then, equation \eqref{eq17} can be simplified as
\begin{equation}\label{eq24}
	\ddot{f} \left(\tau\right)+\dot{f} (\tau)+2\left( \frac{q}{\tau}\right)+\frac{6 q^{2}-2q}{\tau^{2}} f\left(\tau\right)=0\,.
\end{equation}
Due to the cosmological evolution \eqref{eq21}, we can neglect the contribution $\dot{\phi} $. Equation \eqref{eq17} has a solution, which is
\begin{equation}
	f\left( \tau\right)= C_{+} \tau^{\mu_{+}}+C_{-} \tau^{\mu_{-}}\,,
\end{equation}
where $C_{\pm}$ represent constants of integration and
\begin{equation}
	\mu_{\pm} = -q + \frac{1}{2} \pm \frac{\sqrt{-20 q^{2} + 4q+1}}{2}\,.
\end{equation}
Additionally, when the equation above is replaced in equation \eqref{eq15}, $\lambda(\tau)$ is as follows
\begin{equation}\label{eq27}
	\lambda\left(\tau\right) = \mathcal{A}_{+} \tau^{\mu_{+}+6q-2}  + \mathcal{A}_{-} \tau^{\mu_{-} +6q-2}-V\left(\tau \right)\,,
\end{equation}
in which $ \mathcal{A}_{\pm} $ values are obtained as
\begin{equation}
	\mathcal{A}_{\pm}\left( \tau\right) = \left[6q^{2} +6q \mu_{\pm} \right]  C_{\pm} \tau_{0}^{-6 q}\,.
\end{equation}
As expected from the unimodular viewpoint, the Lagrange multiplier $\lambda(\tau)$, represented by equation \eqref{eq27}, is time-varying and represents a time varying the so-called cosmological ``constant". Furthermore, the energy-momentum tensor linked to the equation \eqref{eq7} corresponds to
\begin{equation}\label{eq29}
\left( R_{\mu \nu}-\frac{1}{2} g_{\mu \nu} R\right) =- \frac{1}{f\left(\phi\right)} \left( \left(g_{\mu \nu} \Box - \nabla_{\mu} \nabla_{\nu}\right) f\left(\phi\right) +\left(\frac{\lambda\left( \phi\right)+V\left( \phi\right)}{2}\right) g_{\mu \nu}- \frac{1}{2}g_{\mu\nu} \nabla_{\alpha}\phi \nabla^{\alpha}\phi\right)  \,.
\end{equation}
Therefore, by using equation $G_{\mu \nu}\equiv R_{\mu \nu}-\dfrac{1}{2} g_{\mu \nu} R = \frac{T^{(eff)}_{\mu \nu}}{M_{p}^{2}}$, the effective energy-momentum tensor as
\begin{equation}
	T_{\mu \nu}^{(eff)} \equiv - \frac{M_{P}^{2}}{f(\phi)} \left(\left(g_{\mu \nu} \Box - \nabla_{\mu} \nabla_{\nu}\right) f(\phi)+\left(\frac{\lambda\left( \phi\right)+V\left( \phi\right)}{2}\right) g_{\mu \nu}+ \frac{1}{2}g_{\mu\nu} \nabla_{\alpha}\phi \nabla^{\alpha}\phi\right)\,.
\end{equation}
In this method, we introduce $\rho_{eff}\left(\phi\right)$ and $P_{eff}(\phi)$, which are the effective energy density and the pressure of the Higgs field, respectively; thus, the continuity equation is regarded as
\begin{equation}\label{eq31}
	\dot{\rho}_{eff} + 3 \mathcal{H} \left( \rho_{eff} + P_{eff} \right) =0 \,.
\end{equation}
The field equations \eqref{eq15} and \eqref{eq16} provide $\rho_{eff}\left(\phi\right)$ and $P_{eff}(\phi)$ as follows
\begin{equation}\label{eq32}
	\begin{split}
		& \rho_{eff}\left( \phi\right)  = - \frac{M_{P}^{2}}{f\left( \phi\right)} \left[3 \mathcal{H} \dot{f}\left( \phi\right) a^{6} - \left( \frac{\lambda\left( \phi\right)+V\left( \phi\right)}{2}\right) -\frac{1}{2}\dot{\phi}^{2}a^{6}\right]\,,\\
		& P_{eff}\left( \phi\right)  = -\frac{M_{P}^2}{f(\phi)} \left[ \left( -\ddot{f}(\phi)- 5 \mathcal{H} \dot{f}(\phi) \right) a^{6}+\left( \frac{\lambda\left( \phi\right)+V\left( \phi\right)}{2}\right) -\frac{1}{2}\dot{\phi}^{2} a^{6}\right]\,.
	\end{split}
\end{equation}
From the equations \eqref{eq31} and \eqref{eq32}, we obtain
\begin{equation}\label{eq33}
\begin{split}
		&\ddot{\phi}+3 \mathcal{H} \dot{\phi} \left( \frac{\xi \phi}{f(\phi)-\frac{3}{2}\xi^{2} \phi^{2}}\right)  \left[\frac{3 \xi^{2} \phi^{2}+ 2f(\phi)}{\xi \phi} \right]  +\frac{3}{4}\left( \frac{\xi \phi}{f(\phi)-\frac{3}{2}\xi^{2} \phi^{2}}\right) \dot{\phi}^{2}+ \frac{3}{2}\left( \frac{\xi \phi}{f(\phi)-\frac{3}{2}\xi^{2} \phi^{2}}\right) \xi \dot{\phi}^{2} \\
		&-\left( V\left(\phi\right) + \lambda\left(\phi\right) \right) a^{-6} \left( \frac{\xi \phi}{f(\phi)-\frac{3}{2}\xi^{2} \phi^{2}}\right)+\frac{\left(\lambda^{\prime}\left( \phi\right) +V^{\prime}\left(\phi \right) \right) f(\phi) a^{-6}}{2\xi \phi}\left( \frac{\xi \phi}{f(\phi)-\frac{3}{2}\xi^{2} \phi^{2}}\right)  =0\,,
\end{split}
\end{equation}
where
\begin{equation}\label{eq34}
	\lambda^{\prime}\left( \phi\right) = \frac{d\lambda}{d\phi }\,.
\end{equation}
In these equations, a \textquotedblleft dot\textquotedblright means the differentiation with respect to time parameter $\tau $, and a \textquotedblleft prime\textquotedblright\, indicates differentiation with respect to the scalar field $\phi$. Henceforth, we will not show the explicit dependence of $\lambda $, $ V $, and $ \lambda^{\prime} $, $ V^{\prime} $ to $ \phi $ for the sake of economy. \\

\subsection{Inflation and the slow-roll parameters}
We employ the setup from the \textquotedblleft Hubble Slow-Roll\textquotedblright approximation proposed in Ref. \cite{liddle1994formalizing}. The slow-roll approximations are taken into consideration as \cite{fakir1990improvement}
\begin{equation}\label{eq35}
	\begin{split}
		\left|  \frac{\ddot{\phi}}{\dot{\phi}} \right|  \ll H\,,
		\qquad \left| \frac{\dot{\phi}}{\phi} \right|  \ll H\,,
		\qquad \left| \dot{H}\right| \ll H^{2}\,, \qquad \dot{\phi}^{2}\ll V\,.
	\end{split}
\end{equation}
Then, the conditions in equation \eqref{eq35} are reformulated as
\begin{equation}\label{eq36}
	\frac{d^2 \phi}{d\tau^{2}} \ll -2 \mathcal{H} \frac{d \phi}{d\tau}\,,
	\qquad \frac{d \phi}{d\tau}\ll \mathcal{H} \phi\,,
	\qquad \frac{d\mathcal{H} }{d\tau}\ll -2 \mathcal{H}^{2}\,, \qquad \left( \frac{d\phi}{d\tau}\right)^{2}  \ll V\,.
\end{equation}
The equation \eqref{eq15} may be simplified as
\begin{equation}\label{eq37}
\mathcal{H}^{2} \simeq\frac{2 \left( \lambda+V\right)  a^{-6}}{3 M_{P}^{2} \left( 1+\dfrac{\xi \phi^{2}}{M_{P}^{2}}\right)} \,,	
\end{equation}
and from equation \eqref{eq33} we obtain
\begin{equation}\label{eq38}
3\mathcal{H} \dot{\phi} \simeq \frac{1}{\left( \xi \phi^{2}\left(1+3\xi \right)+M_{P}^{2}\right)}\left( \left( \left( \lambda+V\right)a^{-6}\right) -\frac{\left( \lambda^{\prime}+V^{\prime}\right) \left(1+\frac{\xi\phi^{2}}{M_{P}^{2}}\right) M_{P}^{2}a^{-6}}{4\xi\phi}\right)\,.
\end{equation}
Therefore, the slow-roll indices are defined by \cite{liddle1994formalizing, nojiri2017modified}
\begin{equation}\label{eq39}
\epsilon_{1} = - \frac{\dot H}{H^{2}}\,, \quad
\epsilon_{2} =  \frac{\ddot{\phi}}{H \dot{\phi} }\,, \quad
\epsilon_{3} = \frac{\dot{f}\left(\phi\right) }{2 H f\left(\phi\right) }\,, \quad
\epsilon_{4} =\frac{\dot{E}}{2 H E}\,,
\end{equation}
where
\begin{equation}\label{eq40}
E\equiv f\left(\phi\right)+ \frac{3 \dot{f}^{2}\left(\phi\right)}{2 \kappa^{2} \dot{\phi}^{2}}\,.
\end{equation}
Moreover, the inflationary parameters can be expressed in terms of the slow-roll indices, which have the following form
\begin{equation}\label{eq41}
	n_{s} \simeq 1- 4 \epsilon_{1} - 2 \epsilon_{2} +2\epsilon_{3} -2 \epsilon_{4}\,,
\end{equation}
\begin{equation}\label{eq42}
	r= 8 \kappa^{2} \frac{Q_{s}}{f\left( \phi\right)}\,,
\end{equation}
where $Q_{s}$ is defined as follows
\begin{equation}\label{eq43}
	Q_{s}\equiv \frac{E \dot{\phi^{2}} }{f\left( \phi\right)  H^{2} (1 +\epsilon_{3} )^{2}}\,.
\end{equation}
Note that the expressions \eqref{eq41} and \eqref{eq42} are applicable only when the conditions $\epsilon_{1}, \epsilon_{2}, \epsilon_{3}, \epsilon_{4} \ll 1 $ are fulfilled.
Then, the number of e-folds is written as
\begin{equation}\label{eq44}
	N= \int_{t_{hc}}^{t_{e}} H \, dt = \int_{\phi_{hc}}^{\phi_{e}} \frac{H}{\dot{\phi}} \, d\phi\,,
\end{equation}
where the cosmic time and the scalar field value of the horizon crossing are denoted by ${t_{hc}}$ and $\phi_{hc}$, respectively. Similarly, $t_{e}$ represents the time of inflation's end, and $\phi_{e}$ denotes the corresponding scalar field value at the end of inflation. Thus, the equations given in \eqref{eq39} can be represented as
\begin{equation}
	\epsilon_{1}= -3 - \frac{\dot{\mathcal{H}}}{\mathcal{H}^{2}}\,, \quad
	\epsilon_{2}= 3+\frac{\ddot{\phi}}{\mathcal{H} \dot{\phi}}\,, \quad
	\epsilon_{3}= \frac{\dot{f}\left( \phi\right)}{2 \mathcal{H} f\left( \phi\right) }\,, \quad
	\epsilon_{4}= \frac{\dot{E}}{2 \mathcal{H} E}\,.
\end{equation}
Thus, the function $E$ corresponds to
\begin{equation}
	E= 1 + \left(1 + 6\xi\right) \frac{\xi\phi^{2}}{M_{P}^{2}}\,,
\end{equation}
and
\begin{equation}
		\dot{E}=  \frac{2 \left( 1 + 6 \xi \right)\dot{f}\left( \phi\right)}{M_{P}^{2}}\,.
\end{equation}
To continue, through combination of equations \eqref{eq42} and \eqref{eq43} with the preceding equation, we can determine $r$ as follows
\begin{equation}
	r= \frac{2 M_{P}^{2} E \dot{\phi^{2}} }{f^{2}\left( \phi\right)\mathcal{H}^{2} (1+\epsilon_{3})^{2} }\,.
\end{equation}
According to this theoretical framework, inflation occurs across an expansive range of energy scales, far exceeding those characteristic of the electroweak interaction threshold $ \phi^{2}\ggg\upsilon^{2} $. In essence, as inflation progresses, the potential energy function can be adequately represented by the quartic potential equation, specifically written as $V=\gamma \left(\phi^{2}-\upsilon^{2}\right)^{2}/4\simeq\gamma \phi^{4}/4$, and we define $\lambda= \phi^{2n}/2n$, where $ n $ is an arbitrary power. This simplified model will serve as an appropriate framework for our analysis within this paper. If we take, for example, $n=2 $, the slow-roll conditions \eqref{eq36} allow us to rewrite the equations \eqref{eq37} and \eqref{eq38}, and by taking the necessary derivatives in the following approximate form, we find
\begin{equation}
	\mathcal{H}^{2}\simeq\frac{\left( 1+\gamma\right)  \phi^{2} a^{-6} }{12 \xi} \,,
\end{equation}
\begin{equation}
	\dot{\phi}\simeq -\frac{\left( 1+\gamma\right)a^{-6}}{12 \mathcal{H} \xi^{2}\left( 1+3 \xi \right) }\,,
\end{equation}
\begin{equation}
	\dot{\mathcal{H}}\simeq -\frac{\left( 1+\gamma\right) a^{-6}}{12  \xi^{2} \phi  \left(1+3 \xi \right)}-\frac{\left(1+\gamma \right) \phi^{2}a^{-6}}{4 \xi}\,,
\end{equation}
\begin{equation}
	\ddot{\phi}=\frac{\left( 1+\gamma\right) a^{-6}}{2\xi^{2} \left(1+3 \xi \right)}- \frac{a^{-6}}{12 \xi^{3} \phi^{3} \left(1+3 \xi \right)}-\frac{a^{-6}}{4 \xi^{2} \left(1+3 \xi \right)}\,,
\end{equation}
where we have set $ M_{P}=1$, $1+\gamma \simeq 1$ in the regime where the term $\phi^{2}\gg M^{2}_{P}/\xi$ is dominant. In this case, we have
\begin{equation}
	\begin{split}
		&\epsilon_{1} \simeq\frac{1}{\xi \phi^{3} \left(1+3 \xi \right)}\,, \quad \epsilon_{2} \simeq \frac{1+\left(-9 \gamma -6\right) \xi  \,\phi^{3}}{\xi \phi^{3}\left(1+\gamma \right)}\,, \\
		&\epsilon_{3} \simeq -\frac{1}{\xi\phi^{3} \left(1+3 \xi \right)}\,, \quad \epsilon_{4} \simeq	-\frac{1}{\xi \phi^{3} \left(1+3 \xi \right)}\,,
	\end{split}
\end{equation}
where $\epsilon_{1}\simeq-\epsilon_{3}\simeq-\epsilon_{4}$.
Finally, $ n_{s} $ can be represented in relation to the Higgs field through equation \eqref{eq41} for the action \eqref{eq8} as follows
\begin{equation}
	n_{s}\simeq \frac{- 4.2+ 42.6 \phi^{3} \xi^{2}+\left( 14.2 \phi^{3}- 3.6\right) \xi}{\xi  \left( 1+ 3 \xi \right) \phi^{3}}\,.
\end{equation}
Similarly, $r$ is given by
\begin{equation}
	r\simeq\frac{32+192 \xi}{\xi  \left(6 \phi^{3} \xi^{2}+2 \phi^{3} \xi -1\right)^{2}}\,.
\end{equation}

\begin{figure}[H]
	\centering
	\includegraphics[height= 6cm, width=9cm]{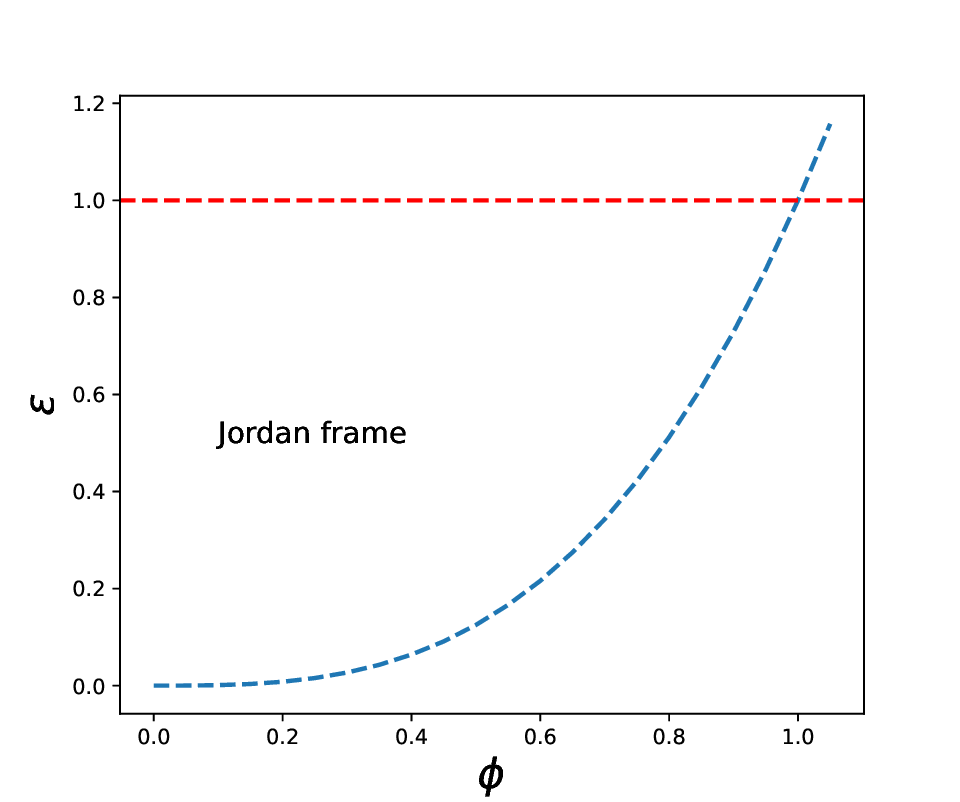}
	\caption{ $\epsilon $ versus $ \phi $ for the Jordan frame.}\label{fig.1}
\end{figure}

It is evident from Figure \eqref{fig.1} that in this model inflation ends gracefully and needs no additional mechanism to be terminated. The end of inflation, that is to say,  $\epsilon_{end}=1$ and $\phi=\phi_{end}$  gives
\begin{equation}
	\phi_{end}=\frac{1}{\left( \xi\left(1+3\xi \right)\right)^{1/3}}\,,
\end{equation}
for our forthcoming purposes. The value of $N$ is determined using equation \eqref{eq43} as follows
\begin{equation}\label{eq58}
	N=\xi\left(1+3 \xi \right)\frac{\phi^{3}}{3} \bigg|_{\phi_{e}}^{\phi_{hc}}\,.
\end{equation}
Considering that $\phi_{e} \ll \phi_{hc}$, we obtain
\begin{equation}
	\phi_{hc}\simeq \left(\frac{3N}{\xi\left(1+3\xi \right)} \right)^{1/3}\,.
\end{equation}
Finally, in our setup $ n_{s} $ and $r$ as two inflation observable are follows
\begin{equation}
	n_{s}\simeq \frac{\left(- 4.2+\frac{ 127.8 N \xi}{1+3 \xi}+\left(\frac{ 42.6 N}{\xi  \left(1+3 \xi \right)}- 3.6\right) \xi \right) \left(1+3 \xi \right)}{3 \left( 1.0+ 3.0 \xi \right) N}\,,
\end{equation}
\begin{equation}
	r\simeq \frac{32+192 \xi}{\xi  \left(\frac{18 N \xi}{1+3 \xi}+\frac{6 N}{1+3 \xi}-1\right)^{2}}\,.
\end{equation}

At this stage, we examine the parameter space of the model through a numerical analysis. Our strategy is to choose appropriate values of $\xi$ from the domain of this quantity based on standard Higgs inflation and then to see what are the inflation observable $n_s$ and $r$ in this setup. On the other hand, the observationally acceptable values of $n_s$ and $r$ restrict the domain of $\xi$ accordingly. We realize that $\xi$ may have three different compatible values ( That is, $\xi= 6 \times 10^{2},  6.01 \times 10^{2}, 6.1 \times 10^{2}$) for $n_s$ and $r$ in the observational domain. So, in Tables \eqref{table-1}-\eqref{table-3}, we compute these parameters for different values of the number of e-folds, $N$.
Table \eqref{table-1} displays that for $ \xi = 6 \times 10^{2} $ and $ N=55 $, the observable are $n_s= 0.9752$ and $r=0.0017$ that are well inside the confidence levels of Planck2018 and BICEP/Keck data. For this value of the parameter $\xi$, other choices of the e-folds number give results out of the observational confidence levels of Planck2018 and BICEP/Keck data. In fact, for other values of $N$ in this case, we find the blue spectrum of the perturbations. But in this case, the obtained values of the parameter $r$ are acceptable observationally and decreasing by increasing $N$. From the following tables we see that the values of $n_s$ and $r$ are so sensitive to the change of the non-minimal coupling parameter, $\xi$.

\begin{table}[H]
	\centering
		\caption{Values on $  n_{s}$ and $ r$ for $  \xi= 6 \times 10^{2} $.}
		\label{table-1}
		\begin{tabular}{c c c c c}
			\hline
			\hline
			& $ N=55 $ & $ N=60$ & $ N=65 $ & $ N=70 $\\
			\toprule
			$ n_{s} $ & $0.9752$ & $1.9590 $ & $ 2.7913$ & $ 3.5048 $\\
			\hline
			$r $ & $0.0017$ & $0.0014$ & $0.0012$ &$ 0.0010 $ \\
			\bottomrule
		\end{tabular}

	\hfill
	\parbox{.45\linewidth}{
		\centering
		\caption{Values on $n_{s}$ and $ r$ for $  \xi= 6.01 \times 10^{2} $}
		\begin{tabular}{c c c c c}
			\hline
			\hline
			& $ N=55 $ & $ N=60$ & $ N=65 $ & $ N=70 $\\
			\toprule
			$ n_{s} $ & $0.9556$ & $1.9410 $ & $ 2.7747$ & $ 3.4894 $\\
			\hline
			$r $ & $0.0017$ & $0.0014$ & $0.0012$ &$ 0.0010 $ \\
			\bottomrule
		\end{tabular}\label{table-2}
		
	}
\hfill
\parbox{.45\linewidth}{
	\centering
	\caption{Values on $n_{s}$ and $ r$ for $  \xi= 6.1 \times 10^{2} $}
	\begin{tabular}{c c c c c}
	\hline
	\hline
	& $ N=55 $ & $ N=60$ & $ N=65 $ & $ N=70 $\\
	\toprule
	$ n_{s} $ & $0.7789$ & $1.7790 $ & $ 2.6252$ & $ 3.3505 $\\
	\hline
	$r $ & $0.0017$ & $0.0014$ & $0.0012$ &$ 0.0010 $ \\
	\bottomrule
\end{tabular}\label{table-3}
	
}
\end{table}

Which frame is physical and appropriate for the calculations? Currently there exists a serious debate surrounding the notion of frames and their role in defining geometric structures from a physical and conceptual perspectives~\cite{Faraoni1999,Capozziello2010,Kamenshchik2015}. Since it calculates the physical distances using the original metric $g_{\mu \nu}$, the Jordan frame seems to be more appropriate when confrontation with observational data is considered. Nevertheless, Einstein frame has its own peculiarities that make its favorable in some other senses. In the next section we switch to Einstein frame to see how is the status of the Higgs inflation in this frame.\\


\section{The Higgs inflation in the Einstein frame}\label{sec-3}

Now we consider a conformal transformation applied to the action \eqref{eq8} in order to switch to the Einstein frame.  We use a Weyl conformal transformation specified at
\begin{equation}\label{eq62}
	{g}_{\mu\nu} \longrightarrow	\tilde{g}_{\mu\nu}=\Omega^{2} g_{\mu \nu}\,,   \qquad
	\sqrt{-g}=\Omega^{-4} \sqrt{-\tilde{g}}\,,
\end{equation}
where the conformal factor is denoted as $\Omega = \Omega\left( \phi\right)$. It is important to note that after inflation, $\Omega = 1$ and consequently the results for observables computed in two frames would be the same. Then, we define $\varphi$, which is established as a novel scalar field, as follows
\begin{equation}\label{eq63}
	\kappa \varphi \equiv \sqrt{\frac{3}{2} }\ln \Omega^{2}\,, \qquad
	\qquad
	\Omega^{2}= e^{\sqrt{2/3} \kappa \varphi} \,.
\end{equation}
The Ricci scalar is represented by
\begin{equation}
	\tilde{R} = \Omega^{-2} \left[ R -6 \Box \left( \ln{\Omega} \right) - 6 \nabla_{a} \left( \ln{\Omega} \right) \nabla^{a} \left( \ln{\Omega} \right) \right]\,.
\end{equation}
Quantities in the Einstein frame are denoted by a tilde. Then, we have the action in what is known as the \textquotedblleft Einstein\textquotedblright (marked with $E$) frame as
\begin{equation}\label{eq65}
\begin{split}
		S_{E}= &\int d^{4}x  \Biggl\{ \sqrt{-\tilde{g}} \Bigg[  \frac{f\left( \phi\right)}{\Omega^{2}} \left( \tilde{R}+3 \tilde{g}^{\mu\nu} \tilde{\nabla}_{\mu} \tilde{\nabla}_{\nu} \ln\Omega^{2}-\frac{3}{2} \tilde{g}^{\mu\nu} \tilde{\nabla}_{\mu} \ln\Omega^{2} \tilde{\nabla}_{\nu} \ln\Omega^{2} \right)\\
	&- \frac{\tilde{g}^{\mu\nu} \partial_{\mu}\phi \partial_{\nu}\phi}{2\Omega^{2}}-\frac{V\left(\phi \right) }{\Omega^{4}} \Bigg] -\tilde{\lambda} \left( e^{-2\sqrt{2/3}\varphi/M_{P}} \sqrt{-\tilde{g}}-1\right) \Biggr\}\,,
\end{split}
\end{equation}
where $\tilde{\lambda} = \lambda_{0}$ (that is to say, it is now a constant) and $\Omega^{2}=2f\left( \phi\right)/M^{2}_{P}$. As before, we assume $f\left(\phi \right)$ to be the Higgs field non-minimal coupling in the following form
\begin{equation}
	f\left(\phi \right)=\frac{M_{P}^{2}}{2} \left( 1+  \frac{\xi \phi^2}{M_{P}^{2}}\right)\,.
\end{equation}
Now, the new conformal transformation is as follows
\begin{equation}\label{eq67}
	\Omega^{2} = \frac{2 f\left( \phi\right) }{M_{P}^{2}} =\left( 1+ \frac{\xi \phi^{2}}{M_{P}^{2}}\right)\,.
\end{equation}
At this point, the action \eqref{eq65} can be expressed as follows
\begin{equation}
		S_{E}=\int d^{4}x \Biggl\{\sqrt{-\tilde{g}} \Bigg[ \frac{M_{P}^{2}}{2}  \tilde{R}-\frac{1}{2} \left( \sqrt{\frac{1+ \left( 1+6 \xi \right) \xi \phi^{2}/ M_{P}^{2}}{\left(1+ \xi \phi^{2}/M_{P}^{2}\right)^{2}}}\right) \tilde{g}^{\mu\nu} \partial_{\mu}\phi \partial_{\nu}\phi -\frac{V\left(\phi\right) }{\Omega^{4}}\Bigg] -\tilde{\lambda}\left(	e^{-2\sqrt{2/3}\varphi/M_{P}} \sqrt{-\tilde{g}} -1\right)\Biggr\}\,,
\end{equation}
where since the total derivative term $\Omega^{2}$ has no impact on the equations, we have omitted it. In this context, the Higgs field's conformal transformation \eqref{eq62} appears to provide a non-minimal kinetic term, so we have
\begin{equation}\label{eq68}
	\frac{d\varphi}{d\phi}\equiv {M_{P}} \sqrt{\frac{f\left( \phi\right)+ 3 \left(f^{\prime}\left( \phi\right) \right)^{2}}{2 f^{2}\left( \phi\right)}} = \sqrt{\frac{1+ \left( 1+6 \xi \right) \xi \phi^{2}/ M_{P}^{2}}{\left(1+ \xi \phi^{2}/M_{P}^{2}\right)^{2}}}\,,
\end{equation}
for the new scalar field $\varphi$. As a consequence, the total action is simply
\begin{equation}\label{eq70}
	S_{E}=\int d^{4}x \Biggl\{\sqrt{-\tilde{g}} \left[\frac{M_{P}^{2}}{2}  \tilde{R}-\frac{1}{2}\tilde{g}^{\mu\nu} \partial_{\mu}\varphi \partial_{\nu}\varphi-U\left(\varphi\right)\right]- \tilde{\lambda}\left(e^{-2\sqrt{2/3}\kappa\varphi} \sqrt{-\tilde{g}} -1\right)\Biggr\}\,,
\end{equation}
where the scalar potential is determined by
\begin{equation}
	U\left(\varphi\right)\equiv \frac{V\left(\phi\right)}{\Omega^{4}}  =\frac{1}{\Omega^{4}}\dfrac{\gamma}{4} \left(\phi^{2} -\upsilon^{2}\right)^{2} \,.
\end{equation}
Here, by taking the derivative of the action \eqref{eq70} with respect to $\tilde{\lambda}$, we find the constraint equation as
\begin{equation}
	\sqrt{- \tilde{g}}=  e^{2\sqrt{2/3}\varphi/M_{P}}\,,
\end{equation}
and consequently, although in the Jordan frame the metric determinant $\tilde{g}_{\mu \nu} $ is fixed, in this frame it is not a constant \cite{saez2016analyzing}. Then, we obtain the equations of motion as
\begin{equation}
	\tilde{R}_{\mu\nu} -\frac{1}{2} \tilde{g}_{\mu \nu} \tilde{R} = \frac{1}{M_{P}^{2}}\left[ \nabla_{\mu}\varphi \nabla_{\nu}\varphi -\frac{1}{2}\tilde{g}^{\mu \nu} \nabla_{\mu}\varphi \nabla_{\nu}\varphi -\tilde{\lambda} \tilde{g}_{\mu \nu}  e^{-2\sqrt{2/3} \varphi/M_{P}}-\tilde{g}_{\mu \nu} U\left(\varphi\right)\right]\,.
\end{equation}
Therefore, by using equation $ \tilde{G}_{\mu \nu} = \kappa^{2} \tilde{T}_{\mu \nu} $, the matter energy-momentum tensor is given as
\begin{equation}
	\tilde{T}_{\mu \nu} = \nabla_{\mu}\varphi \nabla_{\nu}\varphi - \frac{1}{2} \tilde{g}_{\mu \nu} \nabla^{\rho}\varphi \nabla_{\rho}\varphi - \tilde{\lambda} \tilde{g}_{\mu \nu} e^{-2\sqrt{2/3}\varphi/M_{P}}-\tilde{g}_{\mu \nu} U\left(\varphi\right)\,.
\end{equation}
In the Jordan frame, the Lagrange multiplier changes over time, unlike in this frame, where it stays constant ($\tilde{\lambda}=\tilde{\lambda}_{0}$), as we have demonstrated in Ref. \cite{Malekpour:2023lsf}.
In this setup, we redefine the action equation \eqref{eq70} as
\begin{equation}\label{eq75}
	S_{E}= \int d^{4}x \sqrt{-\tilde{g}} \left\lbrace \frac{M_{P}^{2}}{2}\tilde{R} - \frac{1}{2} \tilde{g}^{\mu \nu} \nabla_{\mu}\varphi \nabla_{\nu}\varphi - U_{eff}\left( \varphi\right)\right\rbrace \,,
\end{equation}
where the effective potential is given by
\begin{equation}\label{eq77}
	U_{eff}\left( \varphi\right)= U\left(\varphi\right)+\tilde\lambda_{0} e^{-2\sqrt{2/3}\kappa \varphi}\,.
\end{equation}
Equation \eqref{eq75} looks like the standard GR when the non-minimal coupling is eliminated by the Weyl transformation in the Einstein frame. In this regard, the FRW metric \eqref{eq13} is rewritten as
\begin{equation}
	d\tilde{s}^{2}=\Omega^{2} ds^{2}= -\tilde{a}^{-6} (\tilde{\tau}) d\tilde{\tau}^{2} + \tilde{a}^{2} (\tilde{\tau}) \sum_{i=1}^{3}\left(dx^{i} \right)^{2}  \,.
\end{equation}
We can see that $\tilde{a} \equiv \Omega a = \sqrt{f} a $ and $ d\tilde{\tau} \equiv \Omega d\tau = \sqrt{f} d\tau $.
Therefore, in the Einstein frame, the field equations decrease to
\begin{equation}\label{eq78}
	\begin{split}
		&2 \dot{\tilde{\mathcal{H}}} + 9 \tilde{\mathcal{H}}^{2}= -\frac{1}{M_{P}^{2}} \left[\frac{1}{2} \dot{\varphi}^{2} - U_{eff}\left( \varphi\right) \right]  \,, \\
		& 3 \tilde{\mathcal{H}}^{2} = \frac{1}{M_{P}^{2}}\left[ \frac{1}{2}  \dot{\varphi}^{2} +  U_{eff}\left( \varphi\right)\right]  \,,  \\	
		& \ddot{\varphi} + 3 \mathcal{H} \dot{\varphi}+ U_{eff}^\prime\left( \varphi\right) =0 \,,
	\end{split}
\end{equation}\\

\subsection{Slow-roll field equations}
\begin{figure}[h]
	\centering
	\includegraphics[height= 8cm, width=14cm]{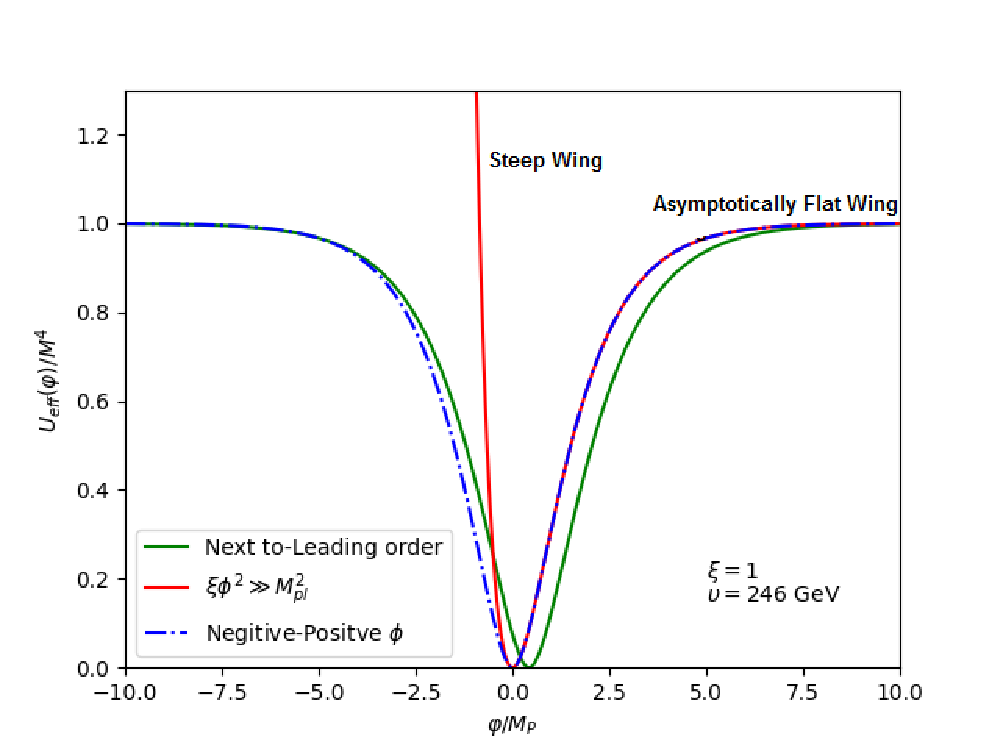}
	\caption{The potential $U_{eff}$ versus the Higgs field $\varphi$ for $ \xi= 1$ and $\upsilon= 246$ GeV. Higgs potential is compared with a leading-order ($\xi \phi^{2}\gg M_{P}^{2}$) equation \eqref{eq86} (red solid line), next-to-leading \eqref{eq87} (green solid line), and the whole field range \eqref{eq89} (blue dotted line) following Refs.~\cite{Garcia-Bellido:2008ycs, martin2016have}.}\label{fig.2}
\end{figure}

In Einstein frame we encounter an effective potential and therefore we can use the \textquotedblleft potential slow-roll approximations\textquotedblright. The inflationary circumstances are identical to those in equation \eqref{eq36}, but with a tilde on all the quantities. If $ \xi \phi^{2} \gg M_{P}^{2} $ and the conditions \eqref{eq36} are satisfied in the Jordan frame, it is simple to demonstrate that these conditions are satisfied in the Einstein frame too. This suggests that inflation happens in both frames under certain conditions. By using equation \eqref{eq36}, the equations of motion and equations of field \eqref{eq78} are as follows
\begin{equation}
	\begin{split}
		& 3 \tilde{\mathcal{H}}^{2} \simeq \frac{1}{M_{P}^{2}}U_{eff}\left( \varphi\right) \,,  \\	
		&3 \mathcal{H} \dot{\varphi} \simeq -U_{eff}^\prime\left( \varphi\right)\,.
  	\end{split}
\end{equation}
In this regard, we have the following potential slow-roll parameters~\cite{liddle1994formalizing}
\begin{equation}\label{eq80}
	\begin{split}
		&	\epsilon = \frac{M_{P}^{2}}{2} \left( \frac{U_{eff}^{\prime}\left(\varphi\right) }{U_{eff}\left(\varphi\right)} \right)^{2}\,, \\
		&	\eta = M_{P}^{2} \left(  \frac{U_{eff}^{\prime\prime}\left( \varphi\right) }{U_{eff}(\varphi)} \right)\,,  \\
	\end{split}
\end{equation}
where in inflation period $\epsilon$ and $\eta$ are small as usual, that is, $|\eta| \ll 1$ and $\epsilon \ll 1$, requiring an equally flat potential. The number of e-folds is given by
\begin{equation}\label{eq81}
	N = \int_{\varphi_{hc}}^{\varphi_{e}} \frac{\tilde{H}}{\dot{\varphi}} \, d\varphi =  \frac{1}{M_{P}^{2}} \int_{\varphi_{e}}^{\varphi_{hc}} \frac{U_{eff}(\varphi)}{U^{\prime}_{eff}(\varphi)} d\varphi \,.
\end{equation}
Now, when the differential equation \eqref{eq68} is precisely integrated, the outcome is as follows
\begin{equation}\label{eq82}
	\frac{\sqrt{\xi}}{M_{P}}\varphi=\sqrt{1+6\xi} \sinh^{-1}\left(\sqrt{1+6\xi }u \right)-\sqrt{6\xi} \sinh^{-1} \left(\sqrt{6\xi} \frac{u}{\sqrt{1+u^{2}}}\right) \,, \quad u\equiv \sqrt{\xi} \phi /M_{P}\,.
\end{equation}
The Higgs inflationary paradigm typically assumes a regime of large couplings, denoted by the parameter $\xi \gg 1$. Therefor, for the large coupling limit, we have $ 1+6\xi\approx 6\xi$ \cite{Garcia-Bellido:2011kqb}, and we can approximate equation \eqref{eq82} by
\begin{equation}\label{eq83}
\frac{\sqrt{\xi}}{M_{P}}\varphi	\approx\sqrt{6\xi} \ln\sqrt{\left( 1+u^{2}\right)}\,.
\end{equation}
Thus, for $\xi \gg 1$, the conformal factor is given by 
\begin{equation}\label{eq84}
	\Omega^{2}=e^{\sqrt{\frac{2}{3}}\frac{\varphi}{M_{P}}}\,.
\end{equation}
Since $\upsilon\ll M_{P}$ implies that $ 1+\frac{\xi \upsilon^{2}}{M^{2}_{P}} \approx 1$, we can safely disregard the $\upsilon$ (vacuum expectation value) for the evolution during preheating and inflation. The Higgs field has an exponentially flat potential for large values of $\xi$ and has the form
\begin{equation}\label{eq85}
	U_{eff}\left(\varphi\right) \simeq \frac{\gamma M_{P}^{4}}{4 \xi^{2}} \left(1-e^{-\sqrt{\frac{2}{3}}\frac{\varphi}{M_{P}}} \right)^{2}-\tilde{\lambda}_{0} e^{-2\sqrt{\frac{2}{3}}\frac{\varphi}{M_{P}}}\,.
\end{equation}
Considering $ M_{P}^{4} \gg  \tilde{\lambda}_{0} e^{-2\sqrt{\frac{2}{3}}\frac{\varphi}{M_{P}}}$, in the right hand side of the equation \eqref{eq85} we can ignore the second term. This observation allows us to simplify our analysis by ignoring the second term, making the problem more manageable. Therefore, a good analytical approximation of the potential can be obtained by rewriting equation \eqref{eq85} as follows
\begin{equation}\label{eq86}
	U_{eff}\left(\varphi\right) \simeq \frac{\gamma M_{P}^{4}}{4 \xi^{2}} \left(1-e^{-\sqrt{\frac{2}{3}}\frac{\varphi}{M_{P}}} \right)^{2}\,,
\end{equation}
which is shown in Figure \eqref{fig.2} (red solid line). It has an asymptotically flat right wing and a steep left wing. In this way, by replacing $\tilde{\lambda}_{0}=0$, the Starobinsky inflation is restored. However, the above potential is approximate, whereas Starobinsky is an exact solution. The two models are not exactly the same, but they do match at \textit{leading order} in $ 1/\xi $ only.

As discussed in Ref. \cite{Martin:2013tda}, within the framework of the Higgs inflation model, equation \eqref{eq86} serves as an approximate representation of the effective potential function, albeit with certain limitations. Specifically, this equation was derived by taking the large limit of $\xi\gg 1 $. The Higgs inflationary paradigm relies on the incorporation of two additional parameters, specifically $\xi $ and $\upsilon$, into its fundamental framework.
In Figure \eqref{fig.2}, we show a visual representation of the expression for $\xi=1 $ and $\upsilon= 246 $ GeV and compare it with the approximation derivation in equation \eqref{eq86} (red solid line).
In this manner, as we'll see later, while $\xi=1$ is unreasonably small, it makes it possible to recognize the difference between the curves; otherwise, it would be impossible to identify. It is possible to verify that equation \eqref{eq86} does, in fact, offer a good approximation to the whole potential at large field values.

Let's perform this approximation at \textit{next-to-leading} order to better evaluate its reliability. By expanding equation \eqref{eq82} at order $\xi^{-1}$, that is, at next-to-leading order compared to equation \eqref{eq86}, we derive an expression for the inflationary potential as
\begin{equation}\label{eq87}
		U_{eff}\left(\varphi\right) \simeq \frac{\gamma M_{P}^{4}}{4 \xi^{2}} \left(1-e^{-\sqrt{\frac{2}{3}}x} \right)^{2}\,, \qquad x=\frac{\varphi/M_{P}}{1+1/(12\xi)}-\sqrt{\frac{3}{2}}\frac{1+\ln(24\xi)}{1+12\xi}\,,
\end{equation}
and for $ \xi\gg 1$, one has $x\simeq \varphi/M_{P}$, so equation \eqref{eq86} is obtained. In Figure \eqref{fig.2}, it is shown with a green solid line, which gives a very good approximation to the whole potential in the inflating zone.

As Figure \eqref{fig.2} illustrates, the approximated Higgs potential, equation \eqref{eq86}, neglects the region $\varphi<0$, where we compare the next-to-leading (green solid line) obtained from equation \eqref{eq87}, with the equation \eqref{eq86} (red solid line).
In the region of positive $\varphi$, both solutions agree extremely well; for $\varphi<0$, however, they diverge significantly. In the region with a negative field, it is not even clear what the conformal transformation is. By using equations \eqref{eq67} and \eqref{eq84}, we have
\begin{equation}\label{eq88}
	\frac{\xi \phi^{2}}{M_{P}^{2}}= \Omega^{2} -1 =e^{\sqrt{\frac{2}{3}}\frac{\varphi}{M_{P}}} -1= \left(1-e^{-\sqrt{\frac{2}{3}}\frac{\varphi}{M_{P}}}  \right) e^{\sqrt{\frac{2}{3}}\frac{\varphi}{M_{P}}}\,,
\end{equation}
that obviously is not consistent; for $\varphi<0$, the equation \eqref{eq88} has a positive definite left-hand side and a negative right-hand side \cite{Garcia-Bellido:2008ycs}. Considering this, we will then employ parametrization
 \begin{equation}\label{eq89}
 		U_{eff}\left(\varphi\right) \simeq M^{4} \left(1-e^{-\sqrt{\frac{2}{3}}|\frac{\varphi}{M_{P}}|} \right)^{2}\,,
 \end{equation}
with
\begin{equation}\label{eq91}
	M^{4}\equiv\frac{\gamma M_{P}^{4}}{4 \xi^{2}}\,.
\end{equation}
The symmetric potential (blue dotted line) \eqref{eq89} shown in Figure \eqref{fig.2} possesses two asymptotically flat wings \cite{Mishra:2019ymr} for the entire span of the field of interest. In contrast to the scenario involving a massive scalar field, the region near $x = 0$ provides a satisfactory level of inflation.
In the framework of Higgs inflation, substituting the value $ M^{4} = 9.6 \times 10^{-11} M_{P} $ (this value was obtained using observational data in Refs. \cite{Mishra:2018dtg, Mishra:2019ymr}) in equation \eqref{eq91}, we find $\xi= 1.62 \times 10^{4} $ for the NMC parameter. Yet inflation holds true in models such as Higgs inflation for values of $\phi\gg M_{P}/\sqrt{\xi}$, with $ \xi\sim 10^{4}$. This observation supports the validity of the Higgs inflationary scenario and highlights the importance of considering the NMC in this framework.

\begin{figure}[H]
	\begin{minipage}[c][1\width]{0.45\textwidth}
		\centering
		\includegraphics[height= 6cm, width=8cm]{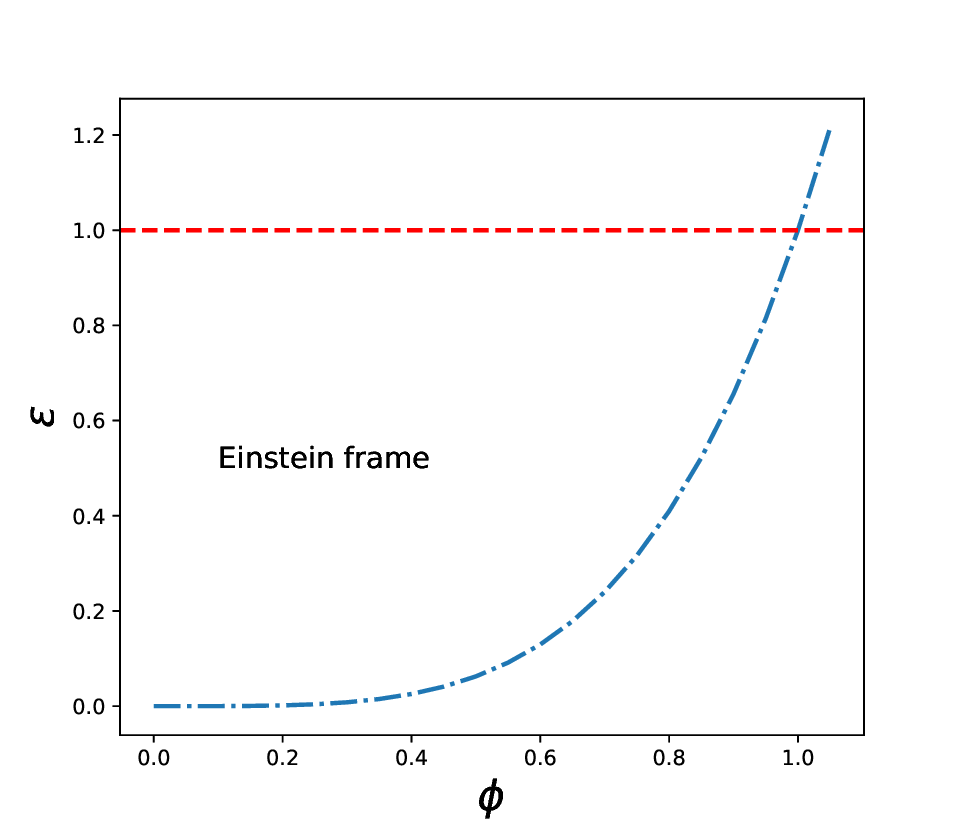}
		\caption{ Evolution $\epsilon $ versus $ \phi $ for the Einstein frame.}
		\label{fig.3}
	\end{minipage}
	\hfill
	\begin{minipage}[c][1\width]{0.45\textwidth}
		\centering
		\includegraphics[height= 6cm, width=9cm]{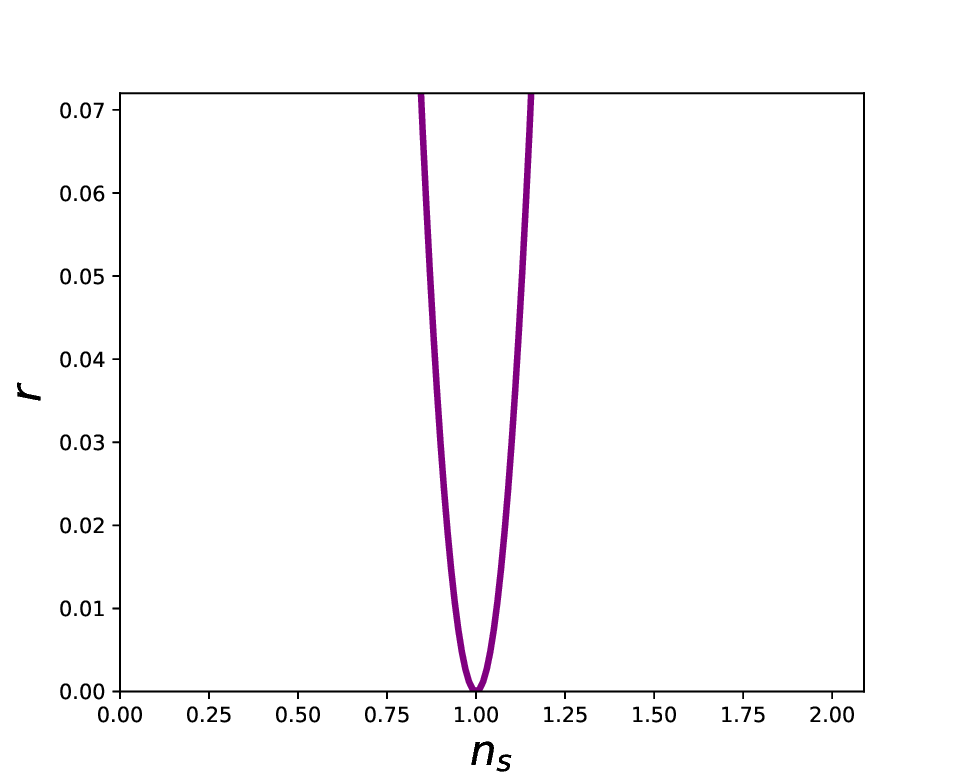}
		\caption{Tensor-to-scalar ratio $r$ versus the scalar spectral index $ n_{s}$ as restricted by the equations \eqref{eq97} and \eqref{eq98}.}
		\label{fig.4}
	\end{minipage}
\end{figure}

Now, by using the potential \eqref{eq89} and equation \eqref{eq68} for the number of e-folds \eqref{eq81} (in the limit of $ \phi^{2}\gg M^{2}_{P}/\xi\gg\upsilon^{2}$), we find
\begin{equation}\label{eq92}
	N\simeq \frac{3 \xi }{4M^{2}_{P}} \left( \phi^{2}_{hc} - \phi^{2}_{e}\right)\,.
\end{equation}
 As the inflation occurs at larger scalar field values, we have used in fact that the $\phi_{hc}\gg\phi_{e} $, therefore we find $\phi_{hc}\simeq2M^{2}_{P}\sqrt{N/3\xi}$. In this case, by using equations \eqref{eq80} we find 
\begin{equation}\label{eq93}
\epsilon \simeq \frac{4 M^{4}_{P}}{3 \xi^{2} \phi^{4}}\simeq\frac{3}{4N^{2}} \,, \qquad \eta\simeq -\frac{4M^{2}_{P}}{3 \xi \phi^{2}}\simeq -\frac{1}{N}\,.	
\end{equation}
Figure \eqref{fig.3} shows the behavior of $\epsilon$ which indicates possibility of graceful exit from inflation phase in this setup. The slow-roll regime ends when $ \epsilon \simeq 1 $, so we have
\begin{equation}\label{eq94}
	\phi_{end}\simeq \frac{M_{P}}{\sqrt{\xi}}\left(\frac{4}{3} \right)^{1/4}\,.
\end{equation}
As a consequence, during inflation we find
\begin{equation}\label{eq95}
	n_{s}\simeq 1-\frac{2}{N}\,, \qquad r \simeq \frac{12}{N^{2}}\,.
\end{equation}
In Figure \eqref{fig.4}, we plot the tensor-to-scalar ratio versus the spectral index ($r$--$n_{s}$) for our model in the Einstein frame. The Planck2018 and BICEP/Keck joint dataset (at $ 95\% $ confidence level) constraint $n_{s}$ and $r$ as  
\begin{equation}\label{eq96}
	n_{s}= 0.9658\pm 0.0038\,,    \qquad   r<0.072\,.
\end{equation}
Hence, in our model for a range of values of the e-foldings number between $55\leq N\leq70$, we obtain
\begin{equation}\label{eq97}
	0.9636\leq n_{s} \leq0.9714\,,
\end{equation}
\begin{equation}\label{eq98}
	0.0039\leq r \leq0.0024\,.
\end{equation}

Thus, as far as $ M_{P}^{4} \gg  \tilde{\lambda}_{0} e^{-2\sqrt{\frac{2}{3}}\frac{\varphi}{M_{P}}}$, the Higgs inflation in unimodular gravity is equally effective and also lead to the result that cosmological constant might becoming dominant at later times, providing a comprehensive explanation of the universe's evolution.

\section{Summary and Conclution}\label{sec-4}
In this paper, we investigated a class of non-minimal Higgs inflation in the framework of unimodular gravity.
We examined a scenario involving cosmological inflation in which the Higgs boson is the inflaton field and exhibits a significant non-minimal coupling to gravity in a unimodular framework. In the first step, we firstly performed our calculations in the Jordan frame and then extended our treatment to the Einstein frame in order to see the status of inflation in these two frames. Essentially, the Higgs field could have a potential to serve as the driving force behind cosmic expansion during the inflationary era. Due to its predictive nature, Higgs inflation enables the determination of various parameters through observations of the CMB. A flat Higgs potential in the large field regime is required for successful inflation to happen, enabling the universe to experience a prolonged period of rapid expansion, during which the universe's density fluctuations are generated and the conditions for the emergence of structure are established. Our proposed model here fulfills more or less these general requirements in addition to some novel achievements such as realization of cosmological constant as a natural outcome of this unimodular setup. It is a common knowledge that unimodular gravity is contained within a subset of general relativity. However, the cosmological constant plays a vital role in this configuration and is typically treated as an integration constant or Lagrange multiplier rather than a fundamental component of the gravitational theory. Through analysis of the results obtained in this study, we have shown that in the presence of the NMC term, it is possible to smoothly exit the inflation phase without needing extra mechanisms in this unimodular non-minimal Higgs inflation. We found also that under slow-roll conditions, our model is able to successfully place the observables within their expected observational ranges (confidence levels), demonstrating its potential to be judged as a successful inflation model. By a full examination of the effective potential in Einstein frame, we explored some important aspects of the inflation model in this frame in connection to the magnitude of the non-minimal coupling parameter, $\xi$. In comparison with observational data from Planck and BICEP/Keck collaborations joint data, these achievements highlight the effectiveness of our approach and underscores the importance of continued development and refinement of models aimed at characterizing the early universe through CMB radiation observations. Through the numerical analysis of the model parameter space, we have now reached the limit on a strong NMC parameter, $\xi\gg1$. In this model, the NMC $\xi$ is determined by the magnitude of scalar perturbations in the light of Planck and BICEP/Keck joint data where a large value for $\xi$, around $10^{2}-10^{4}$, is needed to have successful inflation. Finally, our numerical treatment of the model parameter space specifies that the inflation observables, such as $ n_{s}$ and $r$, are in excellent agreement with Planck2018 and BICEP/Keck data. \\

\end{document}